\documentclass[12pt,a4paper,twocolumn]{narms2}
\usepackage{array}
\usepackage{amsmath}
\usepackage{amssymb}
\usepackage{amsfonts}
\usepackage{graphicx}
\usepackage{aecompl}
\usepackage{ae}
\usepackage{chikako}

\makeatletter

\renewcommand{\section}{\@startsection%
{section}%
{1}%
{0mm}%
{- \baselineskip}%
{0.15\baselineskip}%
{\normalfont\normalsize}}%

\renewcommand{\subsection}{\@startsection
{subsection}%
{2}%
{0mm}%
{-\baselineskip}%
{0.15\baselineskip}%
{\normalfont\normalsize}}%
\makeatother

\newcommand{\Fig}[1]{Fig.~\ref{fig:#1}}
\newcommand{\Tab}[1]{Tab.~\ref{tab:#1}}

\def\bmath#1{\mbox{\boldmath$#1$}} 

\begin{document}

\title{
  Computer simulation of three dimensional shearing of granular materials:\\
  Formation of shear bands
}



\author{
  \large{S\'andor Fazekas}\\
  {\em Theoretical Solid State Research Group of the Hungarian Academy of Sciences}\\
  \large{J\'anos T\"or\"ok}\\
  {\em Department of Chemical Information Technology, BUTE, Hungary}\\
  \large{J\'anos Kert\'esz}\\
  {\em Department of Theoretical Physics, BUTE, Hungary}\\
  \large{Dietrich E. Wolf}\\
  {\em Department of Physics, University Duisburg-Essen, Germany}\\
}

\date{} 

\abstract{
  ABSTRACT:
  We used computer simulations to study spontaneous strain localization
  in granular materials, as a result of symmetry breaking non-homogeneous
  deformations. Axisymmetric triaxial shear tests were simulated by means
  of standard three-dimensional Distinct Element Method (DEM) with
  spherical grains. Carefully prepared dense specimens were compressed
  between two platens and, in order to mimic the experimental conditions,
  stress controlled, (initially) axisymmetric boundary conditions were
  constructed. Strain localization gave rise to visible shear bands,
  previously found experimentally under similar conditions by several
  groups, and different morphologies could be reproduced. We examined 
  the stress-strain relation during the process and found good agreement 
  with experiments. Formation mechanism of shear bands is discussed. 
}



\maketitle
\frenchspacing 

\section{INTRODUCTION}

Spontaneous symmetry breaking in granular materials occurs in
many different forms. Here we focus on strain localization and 
subsequent development of shear bands. Shear bands appear nearly
always if dry granular material is subjected to shear. Its first study
dates back to the nineteenth century and since then it was investigated in
many different geometries and specially designed laboratory tests
(e.g. plane strain, biaxial, and triaxial tests). Here we present 
numerical studies of axisymmetric triaxial tests, the most common 
laboratory tests in Geomechanics.

In a simplified picture, a triaxial test typically consists of a 
cylindrical specimen enclosed between two end platens and surrounded 
by a rubber membrane. An external pressure is applied on the membrane,
either by placing the system into a pressurized fluid, or creating a 
relative vacuum inside the system. The end platens are pressed against
each other in a controlled way, either with constant velocity (strain
control) or with constant force (stress control). The force
resulting on the platens, or the displacement rate of the platens is
recorded, as well as the volume change of the specimen.

The triaxial test is an elementary test, performed to obtain mechanical 
properties of soils. Antifriction devices (lubricated end platens) were 
designed in order to suppress strong heterogeneous responses, such as 
barreling and localization of deformation along failure planes. 
In the past 20 years the study of localization patterns gained more 
attention and strain localization became an important research field,
as experimental
tools as Computed Tomography (CT) became available to study the internal
structure of strained specimens \shortcite{desrues-geo96,batiste-gtj04}. 
Such studies revealed complex 
localization patterns and shear band morphologies depending on the test
conditions.

\section{SIMULATION METHOD} 

We used standard three-dimensional Distinct Element Method
\shortcite{cundall-geo79} (also known as Molecular Dynamics) 
to perform simulations of strain controlled 
triaxial shear tests. As an advantage, contrary to the 
Finite Element Method (FEM) commonly used in simulations of soils, 
the Distinct Element Method (DEM) does not require any macroscopic
\emph{constitutive model}, 
instead it is based on a microscopic 
\emph{contact model} (for a review see 
\shortcite{luding-book04} and references therein). We used the Hertz contact 
model \shortcite{landau-70} with appropriate damping 
\shortcite{brilliantov-pre96},
combined with a frictional spring-dashpot model 
\shortcite{luding-book04}.

\begin{figure}[t!]
\begin{tabular}{cc@{\hspace{3mm}}c}
\includegraphics{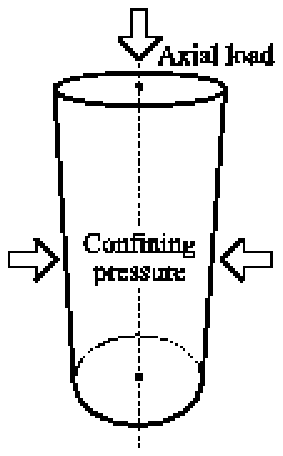} &
\includegraphics[scale=0.5]{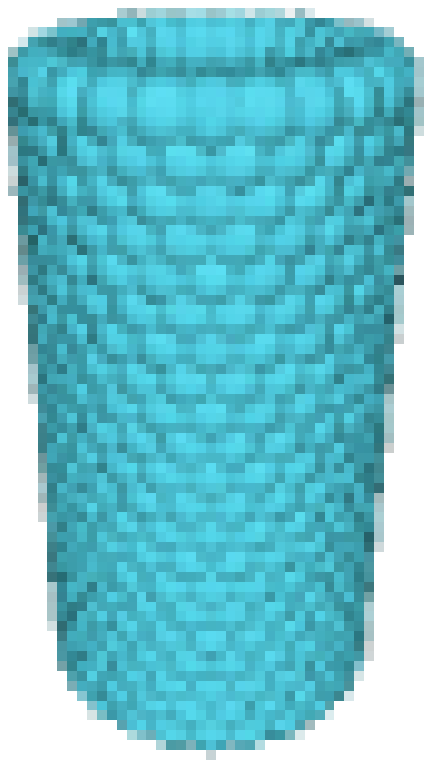} &
\includegraphics[scale=2]{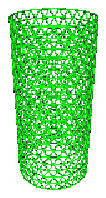} \\
a & b & c
\end{tabular}
\caption{
  Simulation setup. A cylindrical specimen placed between 
  rigid horizontal platens and surrounded by an elastic membrane 
  is subjected to axial load and confining pressure.
}
\label{fig:simmod}
\end{figure}

The simulation setup can be seen on \Fig{simmod}.
An initially axisymmetric system of spherical grains (particles) 
was placed between two horizontal platens. The bottom platen was fixed. 
On the upper platen, having mass $M=10^{-3}\ kg$, 
an axial load was applied (as described later).
The upper platen could not rotate along the vertical axis.
The rotational inertia of the upper platen in tilting
was $I=10^{-7}\ kg\:m^2$.
The particles were surrounded by an ``elastic membrane'' composed of
overlapping spheres having equal diameter $d_m=10^{-3}\ m$ and
mass density ${\rho}_m=0.1 \cdot 10^3\ kg/m^3$, and initially 
forming a triangular lattice (see part (b) of \Fig{simmod}). 
The rotational degree of freedom of 
the ``membrane nodes'' was frozen (i.e. they could not rotate).

The ``membrane nodes'' were interconnected with linear springs 
having zero base length and initial elongation 
$l_0=0.5 \cdot 10^{-3}\ m$ (equal to the initial distance of
neighboring ``membrane nodes''). 
The force $F_s$, acting between two ``membrane nodes'' connected with 
a spring, was calculated as
\begin{equation}
  F_s = {\kappa}_s l_s - {\gamma}_s v_s,
\end{equation}
where ${\kappa}_s=0.5\ N/m$ and ${\gamma}_s=10^{-3}\ N\:s/m$ 
are stiffness and damping
coefficients, $l_s$ is the spring's elongation and $v_s$ is
the relative velocity of the nodes. 
At any time the spring's elongation is equal to the relative distance
of the nodes. The stiffness of the 
springs was chosen such that the particles could not 
escape by passing through the membrane. Additionally a 
confining pressure ${\sigma}_c=0.5 \cdot 10^{3}\ N/m^2$ 
was applied on the membrane, by 
calculating the forces acting on the triangular facets formed by 
neighboring ``membrane nodes'' (see part (c) of \Fig{simmod}). 
The effective
external pressure in this setup is larger than ${\sigma}_c$, 
because the ``membrane springs'' have their own contribution as well,
however, with the used parameters, this contribution 
is small compared to ${\sigma}_c$.

For simplicity we made no difference between particle-particle, 
particle-platen, and particle-membrane contacts.
The normal $F_n$ and tangential ${\bf F}_t$ components
of the contact force were calculated as
\begin{eqnarray}
  F_n &=& 
    {\kappa}_n {\delta}_n^{3/2} 
    - {\gamma}_n {\delta}_n^{1/2} {v_n}, \\
  {\bf F}_t &=&
    {\kappa}_t {\bmath{\delta}}_t
    - {\gamma}_t {\bf v}_t,
\end{eqnarray}
where ${\kappa}_{n}=10^{6}\ N/m^{3/2}$, ${\kappa}_{t}=10^{4}\ N/m$, 
${\gamma}_{n}=1\ N\:s/m^{3/2}$, and ${\gamma}_{t}=1\ N\:s/m$ are the
normal and tangential stiffness and damping coefficients,
${\delta}_n$ and ${\bmath{\delta}}_t$ are normal and tangential 
displacements, and ${v}_n$ and ${\bf v}_t$ are the normal and
tangential relative velocities. 
The normal displacement ${\delta}_n$, the normal velocity ${v}_n$,
and the normal force $F_n$ are
one-dimensional quantities measured along to the normal vector of the 
contact plane, while
the tangential displacement ${\bmath{\delta}}_t$, 
the tangential velocity ${\bf v}_t$, and the tangential
force ${\bf F}_t$ are two-dimensional vectors embedded in the 
contact plane.

Knowing the relative position, the shape, and 
the size of the bodies in contact,
the normal displacement can be calculated directly. Calculating the 
tangential displacement is much more complicated: The tangential 
velocity must be integrated during the lifetime of the contact. 
This integration must be performed in the contact plane. 
In our simulation program the tangential displacement
${\bmath{\delta}}_t$ was implemented as a 3D vector in the
observational space. During integration, this vector was
rotated as the local configuration changed, keeping it always 
in the contact plane.

The Coulomb friction law limits the frictional force to $\mu F_n$, 
where $\mu=0.5$ is the coefficient of friction. To allow for sliding 
contacts, we also limited the length of the tangential displacement
to $\mu F_n / {\kappa}_t$, shortening the displacement vector 
accordingly. This frictional spring-dashpot model implements both 
sliding and static friction.

Our simulations are run at zero gravity.
After calculating the interaction forces, and adding the external load
and confining pressure, the motion of bodies (grains, ``membrane nodes'', 
and upper platen) is calculated by solving numerically the Newton equations 
using a given $\Delta t=10^{-6}\ s$ integration time step. 
The translational motion 
is calculated with Verlet's leap-frog method. The rotational state of 
bodies (given in quaternion representation) is integrated with Euler's 
method. 

In quasi-static processes as the one simulated by us, the vibration
introduced by grain (spring) elasticity is basically undesired noise.
We checked the noise level in our simulations and set the parameters
to keep it low. The inverse of the eigenfrequency of 
all contacts, in both normal and tangential direction, is more 
than one order of magnitude larger than the integration time step.

In the first part of the simulation (preparation phase) a hard 
cylinder touching the internal side of the membrane was introduced, 
and thus the membrane was neglected. The particle system was built
by randomly placing spheres into this cylinder. The maximum 
allowed initial grain overlap was $1\%$. To assure reasonable 
execution time, we started with a sufficiently tall system
(usually $3$ time taller than the final system size).
The particles (grains) were given equal mass density 
${\rho_p}=7.5 \cdot 10^{3}\ kg/m^3$.
The diameter of the spherical grains was taken from a Gaussian 
distribution with mean value 
$\left\langle d \right\rangle=0.9 \cdot 10^{-3}\ m$ 
and standard deviation $\Delta d=0.025 \cdot 10^{-3}\ m$, and
cut at $4 \Delta d$ around the mean value. 

\begin{table}[t!]
\caption{
  Simulation runs. 
}
\vspace{5pt}
\begin{tabular}{>{\footnotesize}l>{\footnotesize}l>{\footnotesize}l}
\hline
      & Upper platen velocity & Upper platen tilting \\
\hline
  (a) & Base                & Enabled \\
  (b) & Base                & Disabled \\
  (c) & Two times faster    & Enabled \\
  (d) & Two times faster    & Disabled \\
\hline
\end{tabular}
\label{tab:simrun}
\end{table}

In the preparation phase the rotational
degree of freedom of the upper platen was frozen and the coefficient 
of friction was set to zero.
Each particle and the upper platen were given a velocity 
proportional to a contraction velocity $v_c=80 \cdot 10^{-2}\ m/s$ 
and their distance from the bottom. 
The system contracted until the upper platen's
velocity in grain-platen collisions decreased to zero. 
At this point an axial load $F_0=200 \cdot 10^{-3}\ N$ was switched 
on, which further contracted the system. After the system 
relaxed, the cylinder was removed, letting the membrane 
and the confining pressure carry the load. The axial load 
$F_0$ and the confining pressure ${\sigma}_c$ were chosen such 
that the system came to equilibrium without barreling.
We prepared one sample having diameter $D=22 \cdot 10^{-3}\ m$ 
and height $H=46 \cdot 10^{-3}\ m$, containing  $N_p=27000$ particles 
and $N_m=14904$ membrane nodes.
The sample's geometry factor $H/D \approx 2$, is similar to the
typical geometry factors used in experiments.

For preparing sphere packings \shortcite{weitz-sci04}
many different methods exist
(e.g. \shortcite{lubachevsky-jsp90,sherwood-jpa97}).
Using the deposition method described above, 
the volume fraction at the end of the preparation phase 
was $f_0=0.643$, which is slightly larger than the random close packing 
value of identical spheres, as expected for an ensemble of spheres with
size dispersion.

After the preparation phase the sample was compressed by moving the
upper platen downward in vertical direction with constant velocity $u_c$
(strain controlled experiment). 
In the executed simulation runs we used two different 
velocities: $u_c=u_0=10^{-2}\ m/s$ (base value), and $u_c=2 u_0$
(two times faster). 
During compression, tilting of the upper platen was either
enabled or disabled. We executed four different
runs (see \Tab{simrun}) starting from the same initial condition.

During the runs
we measured the stress $\sigma$ on the upper platen,
and calculated the stress ratio 
$\sigma/\sigma_0$, where $\sigma_0$ denotes the initial
stress. The identification of the shear bands is a non-trivial
task. One possibility is to generalize the method used in 
two dimensions \shortcite{daudon-pg97}, 
i.e., to calculate the shear intensity around the particles 
from the local deformation tensor. After doing this, we have 
realized that monitoring the rotational state of the particles is
sufficient for the purpose of shear band identificiation, and it leads
to the same results. Our observations are presented in the next
section.

\section{SIMULATION RESULTS} 

\begin{figure}[t!]
\begin{tabular}{c}
\includegraphics{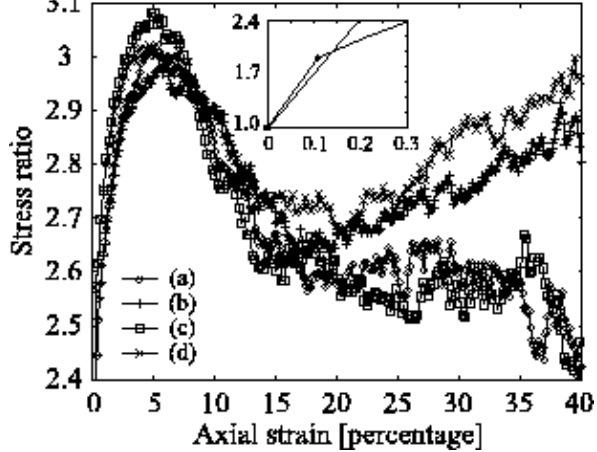} \\
\end{tabular}
\caption{
  Stress-strain relation. The stress ratio 
  ($\sigma/\sigma_0$, where $\sigma_0$ denotes the initial stress)
  measured on the upper platen is shown as function of the axial strain,
  for the executed simulation runs (see \Tab{simrun}).
  (See inset for low stress ratios.)
  For the lower two curves (a, c) tilting of the upper plate was enabled, 
  and for the upper two (b, d) it was disabled. 
}
\label{fig:strainfunc}
\end{figure}

As the axial strain increases, the response of the granular sample 
(the stress ratio) increases until it reaches a peak value 
(see \Fig{strainfunc}). This is a 
basic observation of triaxial shear tests of dense granular specimens
\shortcite{lade-ijss02}. 
Strain localization in granular materials is followed by a decrease in 
load bearing capacity. Dense granular materials dilate during shear.
Due to this, the load bearing particle chains collapse. 
Shear bands occur after peak failure and result in further 
decrease in strength. 

According to \Fig{strainfunc}, up to $15\%$ axial strain there is no 
significant difference in the stress-strain relation measured in the 
different simulation runs, indifferent of the strain rate and tilting 
of the upper platen. However, we observe a change after this point.
At large axial strain values, when tilting of the upper platen was 
disabled, the stress ratio increases again.

\begin{figure}[t!]
\begin{center}
\begin{tabular}{c@{\hspace{10mm}}c}
\includegraphics{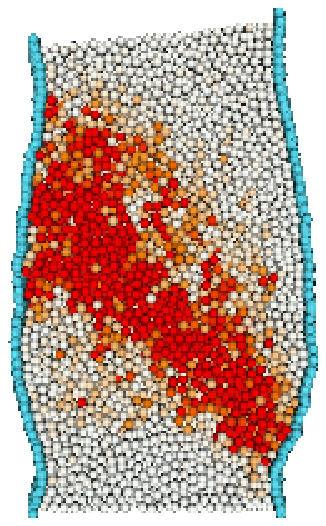} &
\includegraphics{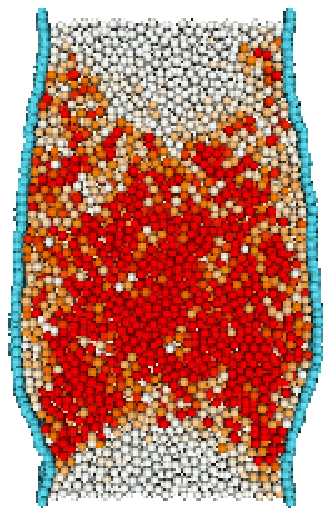} \\
\end{tabular}
\end{center}
\caption{
  Vertical cross sections. 
  The snapshots were taken at the middle of the sample
  at $10\%$ axial strain from the (c) (left) 
  and the (d) (right) simulations. 
  The darkness (red content of color) is proportional to the rotational 
  energy.
}
\label{fig:vercrosec}
\end{figure}

Taking cross sections of the sheared samples, and coloring the grains
according to their rotational state we could visualize the shear bands 
(see \Fig{vercrosec}). On these renderings the larger the rotational 
energy, the more red (or darker in black-and-white version) a particle 
gets. The fact that the rotational state of the grains identifies the 
shear bands, is due to the fact
that the shear bands are characterized not only by 
dilation but also by rotation of the grains, which is known from
both experiments and simulations \shortcite{oda-geo98,herrmann-pha04}.

Another important result is that internal instabilities can develop
into a symmetry-breaking localized deformation along a failure plane
when tilting of the upper platen is enabled, while nontilting platens
act as a stabilizing factor resulting in two axisymmetric conical 
surfaces and complex localization patterns around them
\shortcite{fazekas-tbp05}. This result
is confirmed by similar experiments executed in micro-gravity
\shortcite{batiste-gtj04}, and proves that in the absence of reinforced
axisymmetry, spontaneous symmetry breaking can take place as a result
of internal instabilities \shortcite{desrues-geo96}.

\section{CONCLUSIONS} 

We executed triaxial shear test simulations using DEM. Different 
shear band morphologies known from experiments could be reproduced.
To our knowledge it is the first time that these localization patterns
were reproduced in DEM simulations.
We showed that the shear bands can be identified with the rotational
state of the grains, and symmetry breaking strain localization
can develop if the symmetry is not enforced with nontilting platens.
The agreement of our results with the experimental results is very
good, even if the system size (number of particles) in our 
simulations is much smaller than in experiments.

\section*{ACKNOWLEDGMENTS} 

This research was carried out within the framework of the
``Center for Applied Mathematics and Computational Physics'' of the
BUTE, and it was supported by BMBF, grant HUN 02/011, and Hungarian 
Grant OTKA T035028, F047259.

\bibliography{powderng}
\bibliographystyle{chikako}

\end{document}